\documentclass[12pt]{article}
\usepackage{tabularx}
\usepackage{array}
\usepackage{graphics}
\usepackage{graphicx}
\usepackage{epsfig}
\usepackage{amsmath}
\usepackage{amssymb}
\usepackage{ulem}
\makeatletter

\usepackage{verbatim}
\usepackage{citesort}

\newcommand{\msbar}{{\overline{\rm MS}}}
\newcommand{\ri}{{\rm RI/MOM}}
\newcommand{\bea}{\begin{eqnarray}}
\newcommand{\eea}{\end{eqnarray}}
\newcommand{\beq}{\begin{equation}}
\newcommand{\eeq}{\end{equation}}
\newcommand{\gev}{{\rm GeV}}
\newcommand{\mev}{{\rm MeV}}

\newcommand{\nn}{\nonumber}
\newcommand{\pdir}{p\kern -5.2pt\raise 0.2ex\hbox {/}}
\newcommand{\vdir}{v\kern -5.75pt\raise 0.15ex\hbox {/}}
\newcommand{\kdir}{k\kern -5.75pt\raise 0.15ex\hbox {/}}
\newcommand{\epsdir}{\epsilon\kern -5.0pt\raise 0.15ex\hbox {/}}
\newcommand{\bvdir}{\bar{v}\kern -5.75pt\raise 0.15ex\hbox {/}}
\newcommand{\Ddir}{D\kern -7.75pt\raise 0.20ex\hbox {/}}
\newcommand{\Adir}{A\kern -7.75pt\raise 0.20ex\hbox {/}}
\newcommand{\ldir}{l\kern -5.0pt\raise 0.2ex\hbox{/}}
\newcommand{\varepsdir}{\varepsilon\kern -5.5pt\raise 0.15ex\hbox{/}}

\makeatother

\begin{document}
\thispagestyle{empty} 
\begin{flushright}
\begin{tabular}{l}
{\tt LPT Orsay, 04-17}\\
{\tt RM3-TH/04-5}\\
\end{tabular}
\end{flushright}
\begin{center}
\vskip 1.2cm\par
{\par\centering \Large \bf Estimate of the chiral condensate} \\ 
\vskip 0.2cm\par
{\par\centering \Large \bf in quenched lattice QCD}\\
\vskip 0.9cm\par
{\par\centering   
\sc Damir~Be\'cirevi\'c$^a$ and Vittorio~Lubicz$^b$}
{\par\centering \vskip 0.5 cm\par}
{\sl
$^a$ Laboratoire de Physique Th\'eorique (B\^at 210)~\footnote{Unit\'e mixte de
Recherche du CNRS - UMR 8627.}, Universit\'e de Paris Sud, \\ 
Centre d'Orsay, 91405 Orsay-Cedex, France.
}\\
{\par\centering \vskip 0.25 cm\par}
{\sl
$^b$ Dipartimento di Fisica, Universit\`a di Roma Tre and INFN, Sezione di Roma III, \\ 
Via della Vasca Navale 84, I-00146 Rome, Italy.
}\\

\vskip1.cm

\end{center}

\begin{abstract}
We determine the value of the quark condensate from quenched QCD simulations
on the lattice in two ways: (i) by using the Gell-Mann--Oakes--Renner (GMOR) formula; 
(ii) by comparing the OPE prediction for the Goldstone pole contribution to the pseudoscalar vertex, 
at moderately large momenta. In the $\msbar$ 
scheme at $\mu=2$~GeV, from the GMOR formula we obtain $\langle \bar q q\rangle 
= -\left( 273\pm 19 \ \mev\right)^3$. We show that the value extracted from the 
pseudoscalar vertex, $\langle \bar q q\rangle = - \left( 312\pm 24\ \mev 
\right)^3$, although larger, is consistent with the result obtained from 
the first (standard) method. 
\end{abstract}
\vskip 2.2cm
\setcounter{page}{1}
\setcounter{equation}{0}

\renewcommand{\thefootnote}{\arabic{footnote}}
\vspace*{-1.5cm}

\setcounter{footnote}{1}

\section{Motivation}
\setcounter{equation}{0}
The value of the quark condensate was, and still is, a subject of some 
controversies. It has been experimentally established that in the theory with 
the spontaneous symmetry breaking pattern $SU(2)_L\otimes SU(2)_R \to SU(2)_V$, 
the quark condensate is indeed the order parameter~\cite{leutwyler}. The 
extension to the three flavour case still needs to be clarified (for a recent 
critical discussion see ref.~\cite{stern}). Lattice QCD provides, in principle,
the method for determining the value of the quark condensate and for studying 
its dependence on the number of dynamical quark flavours. Up to now, the 
determination of the quark condensate on the lattice was limited to the quenched
QCD (i.e. with $n_{\rm F}=0$).~\footnote{See ref.~\cite{wittig} for recent
results and the exhaustive list of references.}  Before tackling the theory with $n_{\rm F}=2$ and 
$n_{\rm F}=3$ flavors, one would like to learn as much as possible from the 
quenched theory. For example, one would like to understand if the values of the
chiral condensate obtained by using different methods are consistent among 
themselves.

The standard method relies on the use of the 
GMOR formula, i.e. on the same set of the background 
gauge field configurations one computes both the quark masses 
($m_q$)  and the corresponding pseudoscalar meson masses ($m_P$), and from the 
slope 
\bea\label{1}
m_P^2 = 2 B_0\ m_q\,,
\eea
one gets an estimate of the quark condensate as 
\bea\label{2}
B_0 = - \frac{2}{f^2} \langle \bar q q\rangle\,,
\eea
where $f$ is the pion decay constant in the chiral limit. The renormalization 
scale and scheme dependence of the chiral condensate is just the inverse of the one 
for the quark mass which was discussed in great detail in ref.~\cite{longZ}.

An alternative way for extracting the value of the quark condensate is from the 
study of the amputated pseudoscalar vertex function, where the $\bar q \gamma_5 
q$ operator is inserted at momentum zero.  At moderately large $p^2$ ($p$ being 
the momentum flowing through the legs of the vertex function), one can compare 
the shape of this function with the corresponding expression derived by means of
the operator product expansion (OPE), in which the quark condensate appears in 
the coefficient of the leading power correction.  The lattice estimate based on
this strategy, which is the purpose of this 
letter, has not been presented so far. We show that its value in the continuum limit is fully consistent with 
the standard value, obtained by using the GMOR formula, whose value we updated 
here as well.

\section{Pseudoscalar vertex}
In this section we discuss the relation between the pseudoscalar vertex and the 
quark propagator and study the dependence of these functions on the chiral quark
condensate, which enters their OPE as a leading power correction.

The starting point is to define the quark propagator and the Green function of 
the pseudoscalar density with zero momentum insertion, 
\bea
S(p) = \int dx\ e^{-ipx} \langle    q(x) \bar q(0)  \rangle\,,\quad G_P(p) = 
\int {dx\ dy}\ e^{-ip(x-y)}\langle   q(x) \ \bar q(0) \gamma_5 q(0)\  \bar q(y) 
\rangle\,.
\eea
The amputated vertex function, 
\bea
\Lambda_P(p) = S^{-1}(p) G_P(p) S^{-1}(p)\,,
\eea
is then conveniently projected onto its tree level value 
\bea
\Gamma_P(p) = \frac{1}{12}{\rm Tr}\left[ \gamma_5 \Lambda_P(p) \right]\,,
\eea
where the trace goes over Dirac and color indices so that the factor $1/12$ 
simply provides the normalization to unity. 

If we write the bare (lattice regularized) inverse quark propagator as 
\bea
S^{-1}(p) = \Sigma_1(p^2) \pdir + \Sigma_2(p^2) \,, 
\eea
then the  basic $\ri$ renormalization condition for the quark propagator in the 
chiral limit can be written as~\cite{guido}~\footnote{In practice, we are away 
from the chiral limit, but the renormalization condition applies equally well 
for $m_q^2/p^2\ll 1$~\cite{weinberg}.}
\bea
\left. \frac{1}{Z_q(\mu^2)} \times \frac{1}{12} \frac{{\rm Tr} \left( \pdir 
S^{-1}(p^2)\right)}{p^2}\right|_{p^2=\mu^2} \equiv \left.\frac{\Sigma_1(p^2)}
{Z_q(\mu^2)} \right|_{p^2=\mu^2} = 1\,,
\eea
where $Z_q(\mu)$ is the quark field renormalization ($\widehat S(p,\mu)=Z_q(\mu)
S(p)$). 

By studying the quark propagator at large momenta, one can get an estimate of 
the quark mass value, in the $\ri$ scheme, as
\bea\label{eq:qm1}
m_q^\ri(\mu^2) = \frac{1}{12} {\rm Tr} \left[ \widehat S^{-1}(p,\mu)
\right]_{p^2=\mu^2}=\ \left.\frac{\Sigma_2(p^2)}{\ \Sigma_1(p^2)\ }
\right|_{p^2=\mu^2}\,.
\eea
This estimate has been already discussed in ref.~\cite{bglm}. At lower momenta,
however, this definition of the quark mass suffers from the presence of the long
distance contributions due to the coupling to the Goldstone bosons. 

The effect of the Goldstone boson is more clearly seen by considering the quark 
Ward identity which relates the inverse quark propagator to the amputated
pseudoscalar Green function,
\bea\label{wi1}
\gamma_5 S^{-1}(p^2) +  S^{-1}(p^2) \gamma_5 = 2  Z_A \rho \ \Lambda_P(p^2)\;,
\eea
where $Z_A\rho$ is the quark mass obtained from the hadronic axial Ward identity
on the lattice.~\footnote{Recall that $2 \rho ={\partial_0 \langle\displaystyle{
\sum_{\vec x}} A_0(x) P(0)\rangle / \langle \displaystyle{\sum_{\vec x}} P(x) 
P(0)\rangle }$, with $P = \bar q \gamma_5 q$, $A_0 = \bar q \gamma_0\gamma_5 q$,
and $Z_A\equiv Z_A(g_0^2)$ is the (known) axial current renormalization 
constant.} After multiplying eq.~(\ref{wi1}) by $\gamma_5$ and by taking the 
trace of both sides, we have
\bea\label{wi111}
\Sigma_2(p^2) = Z_A \rho\ \Gamma_P(p^2)\;.
\eea 
For light quark masses and moderately large momenta, the vertex function $\Gamma
_P(p^2)$ is affected by the long distance effects which are due to the presence 
of the Goldstone boson~\cite{alain}, which by means of the LSZ reduction formula
generates the term proportional to the Goldstone boson propagator $1/(q^2 + 
m_\pi^2)$. Since the operator is inserted at zero momentum, $q^2=0$, the vertex 
function in the chiral limit developes a pole $\propto 1/m_\pi^2 \sim 1/\rho$. 
To account for that effect, we expand the vertex function in powers of the quark
mass,
\bea\label{subtr}
\Gamma_P(p^2,\rho) = \Gamma_P^{\rm subtr.}(p^2) + \frac{B(p^2)}{Z_A \rho} 
+ {C}(p^2) \rho \,,
\eea 
where the first term is the subtracted pseudoscalar vertex, from which the 
hadronic (Goldstone boson) contribution $\propto 1/m_P^2 \propto 1/\rho$ is 
subtracted away. The third term is the linear quark mass correction while the higher
order terms in the expansion, as well as the logarithmic quark mass dependence,
are neglected since we deal with light quark masses varying in a short interval.

The renormalization constant of the pseudoscalar density, $Z_P^\ri(\mu)$, is
defined in terms of the subtracted Green function of eq.~(\ref{subtr}) through
the $\ri$ renormalization condition 
\bea\label{rcP}
\left. \frac{Z_P(\mu^2)}{Z_q(\mu^2) } \Gamma_P^{\rm subtr.}(p^2)\right|_{p^2=
\mu^2}  = 1\,.
\eea
As we already discussed in ref.~\cite{longZ}, the value of $Z_P^\ri(\mu)$
obtained from eq.~(\ref{rcP}) is completely consistent with the one obtained by 
applying the method of ref.~\cite{giusti}, which allows one to circumvent the 
second term on the r.h.s. of eq.~(\ref{subtr}).

After inserting eq.~(\ref{subtr}) in~(\ref{wi111}), multiplying both sides by 
$Z_q^{-1}(\mu)$, and accounting for the renormalization condition~(\ref{rcP}), 
we have
\bea\label{eq:qm2}
\left.\frac{\Sigma_2(p^2)}{ \, \Sigma_1(p^2)\,}\right|_{p^2=\mu^2} = 
\underbrace{\frac{Z_A\ \rho}{Z_P(\mu^2)}}_{m^\ri_{\rm AWI}(\mu)} + \frac{B(p^2)}
{Z_q(\mu^2) }\,, 
\eea
where contributions quadratic in the quark mass have been neglected. The first term on
the right-hand side is the usual short distance quark mass, renormalized in the $\ri$
scheme, derived from the axial Ward identity. Equation~\ref{eq:qm2} differs from
eq.~(\ref{eq:qm1}) for the presence of the second term on the r.h.s., which
represents the power suppressed contribution coming from the Goldstone boson.
It has been shown long ago that, at the leading order in the OPE, this term has
the form~\cite{lane}
\bea\label{deraf}
\left.\frac{  B(p^2)}{ \Sigma_1(p^2) }\right|_{\rm OPE} = c(p^2,\mu) 
\frac{\langle \bar q q\rangle (\mu)}{p^2} + {\cal O}\left( 1/p^4\right)\,.
\eea
From this relation we will derive our first estimate of the quark condensate.

The Wilson coefficient, $c(p^2,\mu)$, has been computed at the next-to-leading 
order (NLO) in QCD perturbation theory~\cite{dRP}. In the $\msbar$ scheme, by 
choosing the Landau gauge (in which the lattice calculations are most easily 
made), and after setting $p^2=\mu^2$, one has~\footnote{For completeness, we 
recall the expression for the 2-loop running coupling
\bea
\alpha_S(p)= \frac{4 \pi}{\beta_0 \log(p^2/\Lambda^2_{\rm QCD})} \left(
1 - \frac{\beta_1 \log\log(p^2/\Lambda^2_{\rm QCD})}{ \beta_0^2 \log(p^2/
\Lambda^2_{\rm QCD})}\right);\;\beta_0 = 11-\frac{2}{3} n_{\rm F},\; \beta_1 = 
102-\frac{38}{3} n_{\rm F}.
\eea
} 
\bea
c^\msbar(p^2) = -\frac{4\pi}{3} \alpha_S(p)  \left[ 1 + \left( \frac{99}{4} - 
\frac{10}{9} n_{\rm F} \right) \frac{\alpha_S(p)}{4 \pi} \right]\,.
\eea 
We notice that the radiative corrections are large so that at moderately large 
$p^2$ they must be included in the analysis when extracting the value of the 
condensate from the lattice data. Besides, the inclusion of the radiative 
corrections is also necessary for specifying the renormalization scheme (the 
leading order anomalous dimension of the quark condensate is universal for all
renormalization schemes). To eliminate the scale dependence of the condensate 
one defines the renormalization group invariant (RGI) quark condensate, which 
at NLO in perturbation theory is related to the $\msbar$ one through
\bea\label{rel}
&&\langle \bar q q\rangle^\msbar (p) =
\left(\alpha_S(p)\right)^{-\frac{\gamma_0}{2\beta_0}} \left[
1 - \frac{\gamma_1 \beta_0 - \gamma_0 \beta_1}{2 \beta_0^2} \frac{\alpha_S(p)}{
4\pi} \right] \langle \bar q q\rangle^{\rm RGI}\,,\nn\\
&&\gamma_0=8 ,\,\qquad\gamma_1^\msbar=\frac{4}{3}\left( 101-\frac{10}{3} 
n_{\rm F}\right)\,,
\eea
and thus at $n_{\rm F}=0$, eq.~(\ref{deraf}) becomes
\bea\label{eq:master}
\left.\frac{  B(p^2)}{ \Sigma_1(p^2) } \right|_{\rm OPE} =  
\underbrace{ -\frac{4\pi}{3} \left(\alpha_S(p)\right)^{7/11} \left[ 
1\ +\ \frac{31945}{1452}\ \frac{\alpha_S(p)}{ 4 \pi} \right]}
_{\displaystyle{c^{\rm RGI}(p)}} \frac{\langle \bar q q\rangle^{\rm RGI}}{p^2} 
+ {\cal O}\left( 1/p^4\right)\,,
\eea

\section{Lattice data and extraction of the quark condensate}

We work with the ${\cal O}(a)$ improved Wilson quark action and use the 
data-sets consisting of ${\cal O}(1000)$ independent gauge field configurations, 
obtained at four  different lattice spacings, corresponding to $\beta = 6.0$, $6.2$, $6.4$,and 
$6.45$. More complete information about the data-sets, as well as the improvement coefficients 
with the appropriate list of references can be found in refs.~\cite{longZ,qmass}. Since
we work at four different lattice spacings, we are able to extrapolate to the continuum 
limit. To eliminate the lattice spacing from the results obtained at 
each lattice coupling, we use the ratio $a/r_0$ computed in ref.~\cite{necco}, 
\bea
\left(a/r_0\right)_\beta = \left\{ 0.1863_{6.0},\ 0.1354_{6.2},\ 0.1027_{6.4},
\ 0.0962_{6.45} \right\}\,,
\eea 
so that all our results will be expressed in units of the scale $r_0$. To 
convert into physical units we will use  $r_0 = 0.530(25)$~fm, which corresponds
to $a^{-1}_{\beta=6.0}=2.0(1)$~GeV. We will also need the quenched value of 
$\Lambda_{\rm QCD}$, for which we take $r_0 \Lambda_{\msbar}^{n_{\rm F}=0}=
0.602(48)$~\cite{capitani}.~\footnote{In physical units, $\Lambda_{\msbar}^{n_
{\rm F}=0} = 0.225(20)$~MeV.}

\subsection{$\langle \bar q q \rangle$ from the pseudoscalar vertex }

\begin{figure}
\vspace*{-0.3cm}
\begin{center}
\begin{tabular}{@{\hspace{-0.25cm}}c}
\epsfxsize10.2cm\epsffile{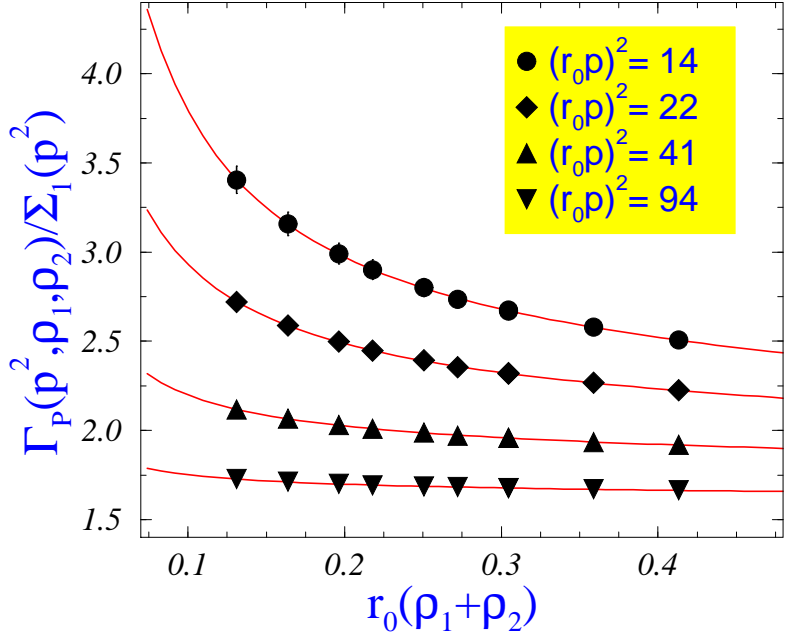}    \\
\end{tabular}
\vspace*{-.1cm}
\caption{\label{fig0}{\footnotesize 
Illustration of the fit to the form~(\ref{fitform00}) at $\beta=6.2$, from which 
we could extract the function $B(p^2)$, needed for the determination of the 
chiral condensate. } }
\end{center}
\end{figure}

In order to determine the chiral condensate from the long distance behavior of
the pseudoscalar vertex,  we first need to extract the function $B(p^2)$. 
That is made by using $10$ different vertex functions, $4$ of which 
are computed with the external legs degenerate in the quark mass, and $6$ 
nondegenerate. With these $10$ points, for each $p^2$, we fit the data to the 
form~(\ref{subtr}), which we rewrite as
\bea\label{fitform00}
\Gamma_P(p^2,\rho_i,\rho_j) = \Gamma_P^{\rm subtr.}(p^2) + \frac{2 B(p^2)}{Z_A 
(\rho_i+\rho_j)} + {C}(p^2)(\rho_i+\rho_j) \,.
\eea
The illustration of this fit is provided in fig.~\ref{fig0} for four values of 
$p^2$. We see that the presence of the Goldstone pole is indeed pronounced at 
moderately large values of $p^2$. 

\begin{figure}
\vspace*{-0.8cm}
\begin{center}
\begin{tabular}{@{\hspace{-0.25cm}}c}
\epsfxsize11.2cm\epsffile{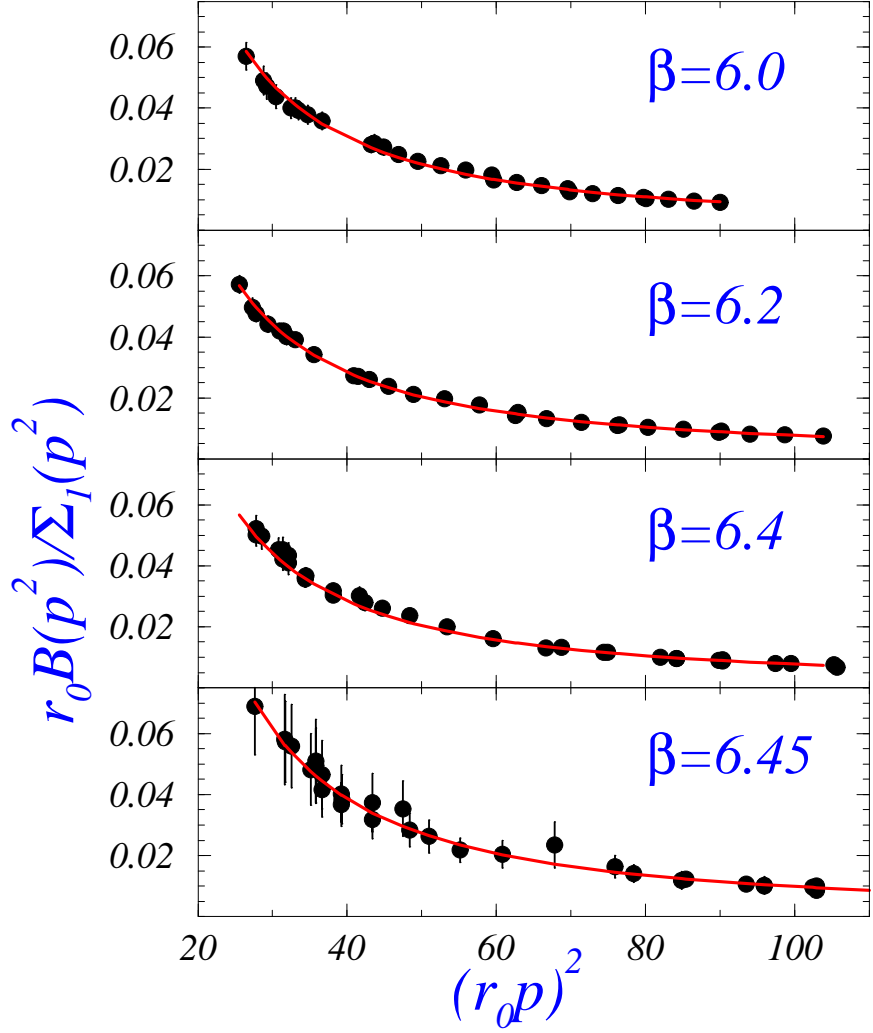}    \\
\end{tabular}
\vspace*{-.1cm}
\caption{\label{fig1}{\footnotesize 
Fit of the lattice data to the form~(\ref{fitform}) for all four lattice 
spacings considered in this letter.}}
\end{center}
\end{figure}

Once we identify the Goldstone contribution to the psudoscalar vertex, we 
perform a number of fits to the form
\bea\label{fitform}
\frac{B(p^2)}{\Sigma_1(p^2) } = c^{\rm RGI}( p ) \frac{\langle \bar q q \rangle
^{\rm RGI}}{p^2} + \frac{\gamma}{p^4} +  \delta +  \lambda  p^2\,,
\eea
where the first term on the r.h.s. is the one that we are interested in (the
coefficient $c^{\rm RGI}$ is defined in eq.~(\ref{eq:master})), the second term 
is the subleading power correction, while the last two terms take into account possible 
contributions of lattice artifacts. To make use of the OPE formula we should aim at working at 
sufficiently large $p^2$ so that higher powers in $1/p^2$ are sufficiently 
suppressed. To do so, we fit the lattice data starting from $p_{\rm cut} 
\approx 2$~GeV, which corresponds to $(r_0 p_{\rm cut})^2 \approx 25$, for which
the radiative correction term in $c^{\rm RGI}(p )$ is below 35\%. 
If we set $\gamma = \delta = \lambda =0$ in (\ref{fitform}), then for all our 
lattices we have $\chi^2/{\rm d.o.f.}> 2$. Therefore one has to let free at 
least one more parameter. The result of such a fit with $\gamma\neq 0$ is 
presented in table~\ref{tab:1} and denoted as fit~I (see also fig.~\ref{fig1} 
for illustration). 
\begin{table}[ht!]
\begin{center} 
\begin{tabular}{|c|c|c|c|c|c|}  
\hline 
{\phantom{\Huge{l}}}\raisebox{-.1cm}{\phantom{\Huge{j}}}
Fit$\quad$ & $\quad\beta\quad$ &  $-r_0 [\langle \bar q q \rangle^{\rm RGI}]
^{1/3}$   & $r_0^5 \gamma$ & $- r_0 \delta \times 10^3$ & 
$-r_0^{-1} \lambda  \times 10^6$  \\ \hline
{\phantom{\Huge{l}}}\raisebox{-.1cm}{\phantom{\Huge{j}}}
& $6.0$ &  $0.71\pm 0.01$ & $\quad 23\pm 3\quad$ & -- & -- \\  
{\phantom{\Huge{l}}}\raisebox{-.1cm}{\phantom{\Huge{j}}}
{\sf I}& $6.2$  & $0.72\pm 0.01$ & $18\pm 1$  & -- & -- \\   
{\phantom{\Huge{l}}}\raisebox{-.1cm}{\phantom{\Huge{j}}}
&$6.4$ &  $0.70\pm 0.02$ & $22\pm 3$  & -- & -- \\  
{\phantom{\Huge{l}}}\raisebox{-.1cm}{\phantom{\Huge{j}}}
&$6.45$ &  $0.75\pm 0.04$ & $31\pm 9$  & -- & -- \\ \hline 
{\phantom{\Huge{l}}}\raisebox{-.1cm}{\phantom{\Huge{j}}}
&$6.0$ & $0.82\pm 0.03$ & $11\pm 5$ & -- & $29\pm 6$ \\  
{\phantom{\Huge{l}}}\raisebox{-.1cm}{\phantom{\Huge{j}}}
{\sf II}&$6.2$  & $0.74\pm 0.02$ & $17\pm 2$  & -- & $3.4\pm 3.7$ \\   
{\phantom{\Huge{l}}}\raisebox{-.1cm}{\phantom{\Huge{j}}}
&$6.4$& $0.73\pm 0.02$ & $20\pm 3$  & -- & $4\pm 2$ \\ 
{\phantom{\Huge{l}}}\raisebox{-.1cm}{\phantom{\Huge{j}}}
&$6.45$& $0.73\pm 0.04$ & $34\pm 12$  & -- & $-4\pm 4$ \\ \hline 
{\phantom{\Huge{l}}}\raisebox{-.1cm}{\phantom{\Huge{j}}}
&$6.0$  & $0.90\pm 0.04$ & $5\pm 6$ & $5\pm 1$ & -- \\  
{\phantom{\Huge{l}}}\raisebox{-.1cm}{\phantom{\Huge{j}}}
{\sf III}&$6.2$  & $0.75\pm 0.03$ & $16\pm 3$  & $1\pm 1$ & -- \\   
{\phantom{\Huge{l}}}\raisebox{-.1cm}{\phantom{\Huge{j}}}
&$6.4$  & $0.75\pm 0.02$ & $18\pm 3$  & $1\pm 1$ & -- \\ 
{\phantom{\Huge{l}}}\raisebox{-.1cm}{\phantom{\Huge{j}}}
&$6.45$  & $0.71\pm 0.05$ & $35\pm 14$  & $-1\pm 1$ & -- \\ \hline 
{\phantom{\Huge{l}}}\raisebox{-.1cm}{\phantom{\Huge{j}}}
&$6.0$  & $0.80\pm 0.02$ & -- & -- & -- \\  
{\phantom{\Huge{l}}}\raisebox{-.1cm}{\phantom{\Huge{j}}}
{\sf IV}&$6.2$ & $0.80\pm 0.01$ & --  & -- & -- \\   
{\phantom{\Huge{l}}}\raisebox{-.1cm}{\phantom{\Huge{j}}}
&$6.4$  & $0.79\pm 0.02$ & --  & -- & -- \\ 
{\phantom{\Huge{l}}}\raisebox{-.1cm}{\phantom{\Huge{j}}}
&$6.45$  & $0.85\pm 0.06$ & --  & -- & -- \\ \hline \hline 
{\phantom{\Huge{l}}}\raisebox{-.1cm}{\phantom{\Huge{j}}}
&$6.0$  & $1.14\pm 0.10$ & -$22\pm 15$ & $27\pm 11$ & $112\pm 73$ \\  
{\phantom{\Huge{l}}}\raisebox{-.1cm}{\phantom{\Huge{j}}}
{\sf V}&$6.2$ & $0.90\pm 0.07$ & $5\pm 6$  & $9\pm 4$ &  $53\pm 27$ \\   
{\phantom{\Huge{l}}}\raisebox{-.1cm}{\phantom{\Huge{j}}}
&$6.4$  & $0.85\pm 0.07$ & $10\pm 4$  & $6\pm 4$ & $24\pm 18$ \\ 
{\phantom{\Huge{l}}}\raisebox{-.1cm}{\phantom{\Huge{j}}}
&$6.45$  & $0.87\pm 0.06$ & $24\pm 11$  & $5\pm 4$ & $27\pm 19$ \\ \hline 
\end{tabular}
\caption{\label{tab:1}
\footnotesize  Details of the fit of the lattice data to the 
form~(\ref{fitform}). Various fit forms (labelled as I, II, III, IV and V) are 
discussed in the text.}
\end{center}
\vspace*{-.3cm}
\end{table}
At fixed lattice spacing, however, the lattice artifacts may be significant. To 
examine their impact on the value of the quark condensate, we repeat the 
fits by including either the term with $p^2$ ($\propto \lambda$) or 
the constant one ($\propto \delta$). Both sets
of results are reported in table~\ref{tab:1}, labelled as fit~II and fit~III, 
respectively. Finally, if we set  $p_{\rm cut}\gtrsim 3$~GeV, the fit with 
$\gamma = \delta = \lambda =0$ gives a satisfactory $\chi^2/{\rm d.o.f.}$ The 
corresponding results are denoted as fit~IV in table~\ref{tab:1}. We also tried
to fit with all the parameters in eq.~(\ref{fitform}) free (fit~V in 
table~\ref{tab:1}).

The remaining step towards the determination of the quark condensate is the 
extrapolation to the continuum limit. Since our action and the renormalization 
constants are ${\cal O}(a)$-improved, we may attempt extrapolating 
quadratically in the lattice spacing, i.e. 
\bea\label{formA}
r_0\left[ \langle \bar q q \rangle^{\rm RGI}\right]^{1/3}_{\beta} = C_0 + C_1 
(a/r_0)^2_{\beta}\,.
\eea
Our result for $C_0$, the chiral condensate in the continuum limit, for all four
fit forms discussed above, are
\bea \label{quadC}
r_0\left[ \langle \bar q q \rangle^{\rm RGI}\right]^{1/3}_{\rm cont.} = \left\{ 
-0.721(23)_{\rm I}, -0.681(28)_{\rm II},-0.672(34)_{\rm III},-0.792(24)_{\rm
IV},,-0.54(15)_{\rm V}
\right\}\,.
\eea
One may argue that terms of ${\cal O}(a)$ may still be present, since the 
function $B(p^2)$ is obtained from the off-shell vertex functions $\Gamma_P(p^2,\rho_i,
\rho_j)$, for which the on-shell ${\cal O}(a)$ improvement 
does not apply. However, the function $B(p^2)$ refers to the chiral limit, and 
terms in $\Gamma_P(p^2,\rho)$ proportional to the quark mass are already taken 
care of in the fit to the form~(\ref{subtr}). In addition, it has been shown in appendix 
of ref.~\cite{longZ} that, for these correlation functions, the ${\cal O}(a)$ 
contribution of operators which are either non gauge-invariant or vanish on-shell by the equation of motion vanish 
in the chiral limit. Therefore, while the ${\cal O}(a)$ effects may affect the functions 
$\Gamma^{\rm subtr.}(p^2)$ and $C(p^2)$ in eq.~(\ref{fitform00}) when away from the chiral limit, 
the function $B(p^2)$ is polluted by the artefacts ${\cal O}(a^2)$ and higher. 
This brings us back to the continuum extrapolation form~(\ref{formA}).

What do we learn from the results~(\ref{quadC}) in the continuum limit? As it 
can be seen from table~\ref{tab:1}, the corrections $\propto 1/p^4$ are large 
and positive for every $\beta$. Their neglect in the fit IV then expectedly lead
to an overestimate of the value for the chiral condensate, as confirmed by the 
last number in eq.~(\ref{quadC}). Fits II and III give quite consistent values 
for the condensate (in the $a\to 0$ limit).  In other words, the quark 
condensate in the continuum limit is very weakly sensitive to the form of the 
artifacts that we include in our fits (constant or $\propto p^2$). The tendency 
of the artifacts, upon their inclusion in the fit, is to lower the value of the 
condensate. The same tendency is observed also in the fit form~V, although with 
larger error bars.

As our final value we will quote the result of the fit I. The difference between
the central value of that and the fits obtained by including the artifacts (II 
and III) is included in the systematic uncertainty. The result of the fit V has
larger errors and is consistent with the results obtained by other fits. 
As we already pointed out, 
the radiative corrections are large and we take them into account when fitting 
the lattice data to eq.~(\ref{fitform}). To account for the systematics induced 
by the omission of higher order corrections in $\alpha_
S(p)$, we will add  $\pm 10$\% of uncertainty (which represents the square of 
the $30$\% effect of the known radiative corrections at $p=3$~GeV). Finally we 
have~\footnote{We remind the reader that $r_0=0.530(25)$~fm, is equivalent to 
$r_0=2.68(13)~\gev^{-1}$.}
\bea\label{res1}
&&\langle \bar q q\rangle^{\rm RGI} = - \left(269\pm 9{}^{+00}_{-18}\pm 12\ 
\mev\right)^3\pm 10\%\,\nn \\ \Leftrightarrow &&\langle \bar q q\rangle^{\rm RGI} = - \left(260\pm 9\pm 9 \pm 12\ 
\mev\right)^3\pm 10\%\,\nn \\ \Rightarrow&&\langle \bar q q\rangle^
\msbar(2~\gev) = - \left( 312\pm 11\pm 11\pm 15\pm 10\ \mev \right)^3,
\eea
where the errors are respectively: statistical, systematics due to the continuum extrapolation, to the uncertainty in $r_0$
and to the uncertainty due to N$^2$LO corrections in the Wilson coefficient
$c^{\rm RGI}(p)$ [see eq.~(\ref{eq:master})]. Notice that in the second line we 
symmetrised the systematic error bars.

Finally, we repeated the entire exercise by using the alternative quark mass 
definition, namely the one derived from the vector Ward identity, $m_q = 
\frac{1}{2}\left(1/\kappa_q - 1/\kappa_{crit}\right)$, instead of the quark mass
$Z_A \rho$, used above. The value we obtain in this way is  barely distinguishable from the 
one we quoted in eq.~(\ref{res1}).~\footnote{More specifically, with $m_q$ 
instead of $\rho$, we get $\langle \bar q q\rangle^\msbar(2~\gev) = - \left( 
313\pm 11\pm 13\pm 15\pm 10\ \mev \right)^3$.}

\subsection{$\langle \bar q q\rangle$ from the GMOR formula}

\begin{table}[ht!]
\begin{center} 
\begin{tabular}{|c|c|c|c|}  
\hline 
{\phantom{\Huge{l}}}\raisebox{-.1cm}{\phantom{\Huge{j}}}
$\beta$ &  $r_0 f$   & $-r_0 \left[\langle \bar q q \rangle^\ri(\mu a=1)\right]
^{1/3}$   & $-r_0 \left[\langle \bar q q \rangle^{\rm RGI}\right]^{1/3}$ \\  
\hline
{\phantom{\Huge{l}}}\raisebox{-.1cm}{\phantom{\Huge{j}}}
$6.0$ & $0.360(7)$ &  $0.709(10)$ & $0.611(9)$ \\  
{\phantom{\Huge{l}}}\raisebox{-.1cm}{\phantom{\Huge{j}}}
$6.2$ & $0.365(10)$ & $0.727(14)$ & $0.612(12)$ \\   
{\phantom{\Huge{l}}}\raisebox{-.1cm}{\phantom{\Huge{j}}}
$6.4$ & $0.358(11)$ & $0.730(16)$ & $0.605(13)$ \\ 
{\phantom{\Huge{l}}}\raisebox{-.1cm}{\phantom{\Huge{j}}}
$6.45$ & $0.365(39)$ & $0.743(50)$ & $0.613(41)$ \\ \hline 
{\phantom{\Huge{l}}}\raisebox{-.1cm}{\phantom{\Huge{j}}}
$\infty$ & $0.362(13)$ & --- & $0.601(25)$ \\ \hline 
\end{tabular}
\caption{\label{tab:2}
\footnotesize  Results of the pseudoscalar decay constant in the chiral limit 
($f$) and the chiral condensate, obtained by means of the GMOR formula (see 
eqs.~(\ref{1}) and (\ref{2}) for all four lattice spacings. We also present the 
results of the linear extrapolation in $a^2$ to the continuum limit ($a\to 
0$).}
\end{center}
\vspace*{-.3cm}
\end{table}

We now repeat the standard exercise of extracting the value of the quark 
condensate by employing the GMOR formula. The values of the pseudoscalar meson 
and the quark masses are all listed in table~2 of ref.~\cite{qmass}. In 
table~\ref{tab:2} of the present letter, we give the results obtained by using 
eqs.~(\ref{1}) and (\ref{2}), where we use for the quark mass the one defined 
via the axial Ward identity ($\rho$). The needed renormalization constants, 
$Z_A$ and $Z_P^\ri(1/a)$, are given in ref.~\cite{longZ}. For completeness, we 
also present the values of the (improved) pseudoscalar meson decay constant in the chiral limit, 
$f$, which is obtained by linearly extrapolating in the quark masses ($f_P=f+ {\rm const.}\times \rho$). To 
convert the quark condensate from the $\ri$ scheme to the RGI form, we use the 
anomalous dimension known up to 4-loops~\cite{chetyrkin}. These 
latter results are then extrapolated to the continuum limit linearly in $a^2$ 
[see eq.~(\ref{formA})]. In physical units, our results read
\bea\label{results2}
\langle \bar q q\rangle^{\rm RGI} = -\left( 224\pm 9\pm 10\ \mev\right)^3\,
\Rightarrow\, \langle \bar q q\rangle^\msbar(2~\gev) = 
-\left( 273\pm 11\pm 15\ \mev\right)^3\,.
\eea
We checked that this value is completely consistent with the alternative 
definition of the quark mass, namely with $m_q = \frac{1}{2}\left(1/\kappa_q - 
1/\kappa_{crit}\right)$, and with $Z_S^\ri(1/a)$ also given in 
ref.~\cite{longZ}.~\footnote{When using the quark mass defined via the vector 
Ward identity instead of the axial one, we get $\langle \bar q q
\rangle^\msbar(2~\gev) = -\left( 268\pm 13\pm 15\ \mev\right)^3$.}
Finally we also note that the above result agrees very well with the QCD sum
rule estimate of ref.~\cite{jamin}, where $\langle \bar q q
\rangle^\msbar(2~\gev) = -\left( 267\pm 16\ \mev\right)^3$ has been
quoted.

\section{Summary and conclusion}

We now briefly summarize our findings and comment on some results reported in
the literature. 

\noindent
{\bf (1)} We update the value of the chiral condensate obtained from  
quenched QCD on the lattice, by using the GMOR formula. After combining the 
errors given in eq.~(\ref{results2}) in the quadrature, we have
\bea
\langle \bar q q\rangle^\msbar_{\rm GMOR}(2~\gev) = 
-\left( 273\pm 19\ \mev\right)^3\,.
\eea 
This result is obtained from simulations performed at four lattice spacings,
by employing non-perturbative renormalization and ${\cal O}(a)$-improvement,
followed by an extrapolation to the continuum limit. 
 
\noindent
{\bf (2)} We compute the quark condensate by using an alternative strategy, 
namely by studying the long distance (Goldstone) part of the pseudoscalar vertex
function. In terms of the OPE, the chiral condensate appears in the coefficient 
of the leading power correction in $1/p^2$. From the calculations at four 
lattice spacings and after extrapolating to the continuum limit we obtain
\bea
\langle \bar q q\rangle^\msbar_{\rm OPE}(2~\gev) = 
-\left( 312\pm 24\ \mev\right)^3\,.
\eea

\noindent
{\bf (3)} From the above results, it seems that the two completely different 
strategies lead to quite a consistent value of the quark condensate. 
To better appreciate this point we rewrite the RGI results in units of 
$r_0$, i.e.,
\bea
-r_0 [\langle \bar q q\rangle^{\rm RGI}]^{1/3} 
= \{ 0.701(23)(20)(23)_{\rm OPE} , 0.601(25)_{\rm GMOR} \} \,.
\eea
The method based on using OPE is less reliable since 
radiative and further power corrections are large. 
Even if we combine the errors in quadrature the agreement would 
be at the $2\sigma$-level, which is far from what has been  
claimed in ref.~\cite{alain}, where the OPE and GMOR results 
were argued to differ by a factor of $3$.

Before closing this letter, we should explain why our conclusion is 
qualitatively different from the one reported in ref.~\cite{alain}. 
The first difference is that in eq.~(\ref{subtr}), besides the 
Goldstone term ($\propto 1/\rho$) we also allow for the presence of the term 
linear in quark mass. Such a term could not be studied in ref.~\cite{alain} 
since only three quark masses were considered. The net effect of this modification 
is that the function $B(p^2)$ becomes smaller. Secondly, in the OPE, we allow 
for the presence of the term $\propto 1/p^4$, and we find that, for moderately large 
momenta, this (subleading) power correction is not negligible, while in ref.~\cite{alain} 
such a term was not found. 
Finally, we also accounted for the terms that are due to the lattice artifacts 
[see eq.~(\ref{formA})], which further reduce the value of the condensate in the
continuum limit. Such effects were not studied in ref.~\cite{alain} (where they only 
considered the data produced at $\beta=6.0$). Notice also that the reference 
value of the chiral condensate considered in ref.~\cite{alain}, was $20$\% 
smaller than the one we obtain here by using the GMOR formula.

\section*{Acknowledgment}
We wish to thank our collaborators of ref.~\cite{longZ}, and especially Alain 
Le Yaouanc, for many useful discussions.

\end{document}